\documentclass[trackchanges,twocolumn]{aastex631}

\usepackage{amsmath}

\begin{document}

\title{Cosmic-Ray Mass Composition around the Knee via Principal Component Analysis}

\author{Nicusor Arsene}
\affiliation{Institute of Space Science - INFLPR Subsidiary, 077125 Magurele, Romania}
\email{nicusor.arsene@spacescience.ro}

\begin{abstract}

In this paper, we apply Principal Component Analysis (PCA) to experimental data recorded by the KASCADE experiment to reconstruct the mass composition of cosmic rays around the \textit{knee} region. A set of four extensive air shower parameters sensitive to the primary particle mass ($LCm$, $N_{\mu}$, $N_{e}$, and lateral shower $age$) was considered, whose coordinates were transformed into a new orthogonal basis that maximally captures the data variance. Based on the experimental distributions of the first two principal components (PCA0 vs.\ PCA1) and full Monte Carlo simulations of the KASCADE array considering five types of primary particles (p, He, C, Si, and Fe) and three hadronic interaction models (EPOS-LHC, QGSjet-II-04, and SIBYLL~2.3d), we obtained the evolution of the abundance of each primary species as a function of energy, as well as the evolution of the mean logarithmic mass with energy. We found that the reconstruction of the mass composition resulting from this comprehensive analysis significantly reduces dependence on the hadronic interaction model used in the simulation process, even though the initial input parameters are model-dependent. Moreover, the results support the idea that around the \textit{knee} region, the abundance of the light component (protons) decreases, while the heavy component shows a slight increase. The evolution of $\langle \ln (A) \rangle$ as a function of energy derived from this analysis shows excellent agreement with recent results from the LHAASO--KM2A experiment and aligns very well with the predictions of the data-driven GSF model.

\end{abstract}


\section{Introduction} 

The origin and acceleration mechanisms of cosmic rays (CRs) are not yet fully identified or understood, although significant progress has been made in these areas in recent years.
Based on measurements performed by multiple CRs experiments \citet{KANG20244403}, it has been established that the CRs flux in the energy range $10^{10} - 10^{20}$ eV can be approximated by a power-law function $dN/dE \sim E^{\gamma}$ \citet{PhysRevLett.125.121106, 2023APh...15102864A, PhysRevLett.132.131002, 2013arXiv1306.6283T, Alfaro:2024qem}. 
This energy spectrum exhibits several highly significant features that could provide insight into the mechanisms by which CRs are accelerated by various astronomical objects, as well as into their propagation through the Galactic and extragalactic medium: a steepening at $\sim 4 \times 10^{15}\,\mathrm{eV}$ (\textit{knee}) \citet{2005APh....24....1A, 2003APh....19..447T, Nagano:1984db, Ogio:2004sc, Fowler:2000si, EAS-TOP:1998aye, PhysRevLett.132.131002, TIBETIII:2008qon}, another at $\sim 8 \times 10^{16}\,\mathrm{eV}$ (\textit{second knee}) \citet{PhysRevLett.107.171104, PhysRevD.100.082002, 2018ApJ...865...74A}, and a flattening at $\sim 5 \times 10^{18}\,\mathrm{eV}$ (\textit{ankle}) \citet{2023APh...15102864A,2023JCAP...05..024A}, reflecting changes in the spectral index $\gamma$.

It was widely accepted that ultra high energy cosmic rays (UHECRs) ($E > 10$ EeV), being less affected by magnetic field deflections, could more directly point back to their acceleration sources. However, recent results from the Telescope Array experiment show that the arrival direction of the highest-energy recorded event indicates no obvious source galaxy \citet{2023Sci...382..903T}. However, the issue remains unresolved and already opens up new scenarios to explain the origin of these extreme energy CRs: either the magnetic fields involved are much stronger than currently expected, or we may be facing an yet-unknown aspect of particle physics.

Significant progress has been made in identifying sources of Galactic CRs, particularly through the detection of sub-PeV diffuse gamma rays from the Galactic disk \citet{PhysRevLett.126.141101, PhysRevLett.134.081002}, as well as ultra-high-energy gamma-ray sources that point to promising PeVatron candidates \citet{2021Natur.594...33C, 2024ApJS..271...25C}.
In this hadronic emission scenario, cosmic ray protons interact with the interstellar medium, producing neutral pions ($\pi^{0}$) which subsequently decay into gamma rays. These findings offers strong evidence for the presence of "PeVatrons" in our Galaxy capable of accelerating cosmic rays well beyond the \textit{knee}, up to several PeV, approaching 10 PeV \citet{LHAASOCOLLABORATION2024449, PhysRevLett.126.141101, PhysRevLett.134.081002}.

A deeper understanding of the origin of sub-PeV Galactic gamma-ray emission demands thorough spectral analysis of individual sources, along with a precise evaluation of the diffuse background contribution to their measured fluxes \citet{Kato:2025gva}. This also requires a good understanding of the mass composition of CRs around the \textit{knee} $\sim 4$ PeV and their propagation process in the Galaxy, in order to more precisely estimate the contribution of sub-PeV diffuse gamma rays to the total flux \citet{thelhaasocollaboration2025identificationprecisespectralmeasurement,PhysRevD.98.043003, 2023A&A...672A..58D, 2025arXiv250218268D}.

In this context, we reassess the mass composition of CRs around the \textit{knee} through a multivariate analysis of data recorded by the KASCADE experiment. 
We investigate the correlation between the $LCm$ parameter \citet{Conceicao:2022lkc} - sensitive to the nature of the primary particle and independent of the hadronic interaction model \citet{2023JCAP...09..020A}, and three other extensive air shower (EAS) parameters commonly used in previous analyses of CRs mass composition. We reconstruct the fractions of different primary species as a function of primary energy in the $\lg(E/\rm{eV}) = [15 - 16]$ range and the evolution of $\langle \ln (A) \rangle$ with energy.

The paper is organized as follows: Section \ref{sec:pca} provides a description of the Principal Component Analysis (PCA) procedure; Section \ref{sec:kascade} gives a brief overview of the KASCADE experiment and the methods used to obtain data and simulations; Section \ref{sec:eas} describes the four EAS parameters sensitive to the nature of the primary particle; Section \ref{sec:mass_comp} presents the reconstruction of individual fractions of different species and mean logarithmic mass $\langle \ln A \rangle$ as a function of energy and compare the obtained results with those recently reported by the LHAASO$-$KM2A experiment, as well as with various data-driven and astrophysical models that describe the evolution of different species of primary particles as a function of energy around the \textit{knee}; and Section \ref{sec:conclusions} concludes the paper.

\section{Principal Component Analysis} \label{sec:pca}

Principal Component Analysis (PCA) involves transforming the initial set of variables through a linear operation that reorients the coordinate system. This is done using an orthogonal matrix, effectively rotating the original space to align with new axes that better highlight underlying structures and help reduce the number of relevant dimensions. In simple terms, the PCA method defines a new orthogonal basis that optimally captures the variance in the data, thereby enhancing the separation between observations \citet{jolliffe2002principal}.
We will briefly describe how the PCA method used in this study works, as it was originally described and implemented by \citet{Holm} in the ROOT framework \citet{BRUN199781}.

Assuming we have $M$ types of primary particles, each characterized by a set of $P$ observables $x_0, x_1, \ldots, x_{P-1}$. Each type of primary particle is a vector in the $P$-dimensional \textit{pattern} space 
\begin{equation}
\mathbf{x}^{(i)} = \left[\begin{array}{c}
x_0^{(i)} \\
x_1^{(i)} \\
\vdots \\
x_{P-1}^{(i)}
\end{array}\right], \quad i = 1, \ldots, M,
\end{equation} 
where $x_n^{(i)}$ represents the value of the $n$-th variable for the $i$-th observation.
The first step involves centering the data by subtracting the sample mean from each observation
\begin{equation}
\bar{\mathbf{x}} = \frac{1}{M} \sum_{i=1}^M \mathbf{x}^{(i)}, \quad
\mathbf{y}^{(i)} = \mathbf{x}^{(i)} - \bar{\mathbf{x}}
\end{equation} where $y^{(i)}$ denote the centered observation vectors.

The sample covariance matrix is computed as:
\begin{equation}
\mathbf{C} = \frac{1}{M} \sum_{i=1}^M \mathbf{y}^{(i)} {\mathbf{y}^{(i)}}^\top = \mathbb{E}\left[\mathbf{y} \mathbf{y}^\top\right],
\end{equation}
where $\mathbb{E}$ denotes the average over all $M$ types of primary particles.
This covariance matrix is symmetric, real, and positive definite, and thus has a full set of orthonormal eigenvectors and non-negative eigenvalues.
The eigenvalues $\lambda_0 \geq \lambda_1 \geq \cdots \geq \lambda_{P-1} \geq 0$
and the corresponding orthonormal eigenvectors $\mathbf{e}_0,\, \mathbf{e}_1,\, \ldots,\, \mathbf{e}_{P-1}$ of the covariance matrix \(\mathbf{C}\) are computed via standard methods:
\begin{equation}
\mathbf{C} \mathbf{e}_n = \lambda_n \mathbf{e}_n, \quad n = 0, \ldots, P-1.
\end{equation}
These eigenvectors define a new orthonormal basis in which the data can be expressed.
The centered data vectors $\mathbf{y}^{(i)}$ are aproximated using the first $N$ principal components:
\begin{equation}
\mathbf{y}^{(i)} \approx \sum_{n=0}^{N-1} a_{i,n} \mathbf{e}_n.
\end{equation}
The projection from the \textit{pattern} space to the \textit{feature} space minimizes the error
\begin{equation}
E_N = \left\langle \left\| \mathbf{y}^{(i)} - \sum_{n=0}^{N-1} a_{i,n} \mathbf{e}_n \right\|^2 \right\rangle.
\end{equation}
Using the condition of orthonormality for $\mathbf{e}_n$ and $a_{i,n} = (\mathbf{e}_n)^\top \mathbf{y}^{(i)}$ the error becomes:
\begin{equation}
E_N = \sum_{n=N}^{P-1} \lambda_n.
\end{equation}
Therefore, selecting the eigenvectors associated with the largest $N$ eigenvalues leads to the smallest approximation error.
The PCA transformation matrix is built from the eigenvectors:
\begin{equation}
\mathsf{T} = \begin{bmatrix}
\mathbf{e}_0 & \mathbf{e}_1 & \cdots & \mathbf{e}_{P-1}
\end{bmatrix},
\end{equation}
and the  projection of an original (centered) vector $\mathbf{y}^{(i)}$ onto the \textit{feature} space is:
\begin{equation} \label{eq:proj}
\mathbf{z}^{(i)} = \mathsf{T}^\top \mathbf{y}^{(i)}.
\end{equation}
By keeping only the first $N$ columns of $\mathsf{T}$, we reduce the dimensionality from
$P$ to $N$ while preserving most of the variance in the data.

\section{KASCADE data and simulations} \label{sec:kascade}

The KASCADE experiment, located in Karlsruhe, Germany, at an altitude of 110 meters above sea level, was dedicated to detecting CRs with energies in the range of $\lg(E/\text{eV}) = [14 - 17]$. The detector array covered an area of $200 \times 200$ m$^{2}$, consisting of 252 detection stations arranged in a rectangular grid with a spacing of 13 meters.
Each station was equipped with both shielded and unshielded detectors, enabling the simultaneous recording of the electromagnetic and muonic components of extensive air showers. The charged particles were detected using liquid scintillation counters placed above the shielding, while the muonic component was measured using plastic scintillators with an area of 3.2 m$^{2}$, located beneath absorbing layers of lead and iron \citet{ANTONI2003490}.
All experimental data collected throughout the entire operational period by the KASCADE collaboration have been made publicly available through the KCDC database \citet{2018EPJC...78..741H}, along with complete sets of Monte Carlo (MC) simulations of the detector array\footnote{https://kcdc.iap.kit.edu}.

The MC simulation process of the entire KASCADE array involved simulating EASs using the CORSIKA code \citet{1998cmcc.book.....H}, while the signal/energy deposited in the detectors was modeled using the CRES package based on GEANT3 \citet{Brun:1987ma}.
Based on the detector response, key EAS parameters were extracted using the KRETA package, including the primary energy, lateral distribution function (LDF), number of muons, number of electrons, \textit{age} parameter, arrival direction, etc.
Both the reconstruction of experimental data and that of simulated data use exactly the same reconstruction procedures.

In this analysis, we considered a set of MC simulations that includes five types of primary particles, namely protons, He, C, Si, and Fe, with energies in the range of $\lg(E/\text{eV}) = [15 - 16]$, following a flux trend with a spectral index of $\gamma = -2.7$.
The zenith angle range was restricted to $\theta = [0^\circ - 20^\circ]$ in order to avoid introducing bias in the $LCm$ parameter (which quantifies the non-uniformity of the signal induced by secondary particles at a given distance around the shower axis), while still ensuring sufficient statistics for the set of MC simulations. The distribution of azimuthal angles is isotropic within the range $\phi = [0^\circ - 360^\circ]$. 
The high-energy hadronic interaction models considered in the EAS simulation process were EPOS-LHC \citet{Pierog:2013ria}, QGSjet-II-04 \citet{Ostapchenko:2004ss}, and SIBYLL 2.3d \citet{Riehn:2019jet}, while low-energy hadronic interactions were modeled using FLUKA \citet{Ferrari:2005zk}. Based on these criteria, the resulting simulation dataset yields a statistics of $\mathcal{O}(10^3 - 10^4)$ events per primary species, per interaction model, and per energy bin of width 0.2 in $\lg(E/\text{eV})$. In the following section, we describe the four EAS observables that are sensitive to the nature of the primary particle, as well as the way they were reconstructed from the KASCADE experimental data.

\section{EAS Observables} \label{sec:eas}

Several EAS observables have been used over time to reconstruct the mass composition of primary cosmic rays (see e.g. \citet{Hoerandel:2002yg} and references therein) and more recently \citet{PhysRevLett.132.131002, thelhaasocollaboration2025identificationprecisespectralmeasurement}. In the energy range $\lg(E/\text{eV}) = [15 - 16]$ eV, the most effective observables have proven to be the number of muons ($N_\mu$) and the number of electrons ($N_e$) from EASs at ground level. The major drawback of these observables is their strong dependence on the high-energy hadronic interaction model considered in the simulation process.

In this study, we combined four EAS observables with the aim of extracting the mass composition from a more comprehensive perspective, namely: $LCm$, $N_\mu$, $N_e$ and the $Age$ parameter (lateral shape parameter), using the KASCADE data. All four of these observables have been shown to be sensitive to the nature of the primary particle.

The $LCm$ parameter, originally introduced as a gamma/hadron discriminator in \citet{Conceicao:2022lkc}, and later analyzed in more detail in various experimental configurations \citet{Bakalova:2023avj, Conceicao:2023xfb, Bakalova:2023efj}, has proven to be an excellent discriminator for mass composition studies when used in detector arrays with sufficiently high density like KASCADE \citet{2023JCAP...09..020A}. This parameter quantifies the non-uniformity of the signal induced in the detectors at a given distance from the shower axis in vertical EASs, and is defined as $LCm = \log(C_k)$ where:
\begin{equation}
C_{k} =\frac{2}{n_{k}(n_{k}-1)} 
\frac{1}{\left<S_{k}\right>}\sum_{i=1}^{n_{k}-1}\sum_{j=i+1}^{n_{k}}(S_{ik}-S_{jk})^{2}.   
\label{eq:CK}
\end{equation}
Here, $n_k$ denotes the total number of detectors in ring $k$, and $\left<S_k\right>$ is the average signal recorded by those detectors. The quantities $S_{ik}$ and $S_{jk}$ correspond to the signals measured by detectors $i$ and $j$ within the same ring. The prefactor $\frac{2}{n_k(n_k - 1)}$ represents the inverse of the number of two-combinations for $n_k$ detectors, $\binom{n_k}{2}$. In our analysis, the signals $S_{ik}$ considered in Equation \ref{eq:CK} represent the energy deposited by the electromagnetic component in the liquid scintillators of the KASCADE array.
At the same energy, $LCm$ values are higher for proton-induced showers compared to iron-induced ones, due to the significantly larger fluctuations in the altitudes at which the primary interactions occur.
\begin{figure*}[ht!]
\centering
\includegraphics[width=0.49\textwidth]{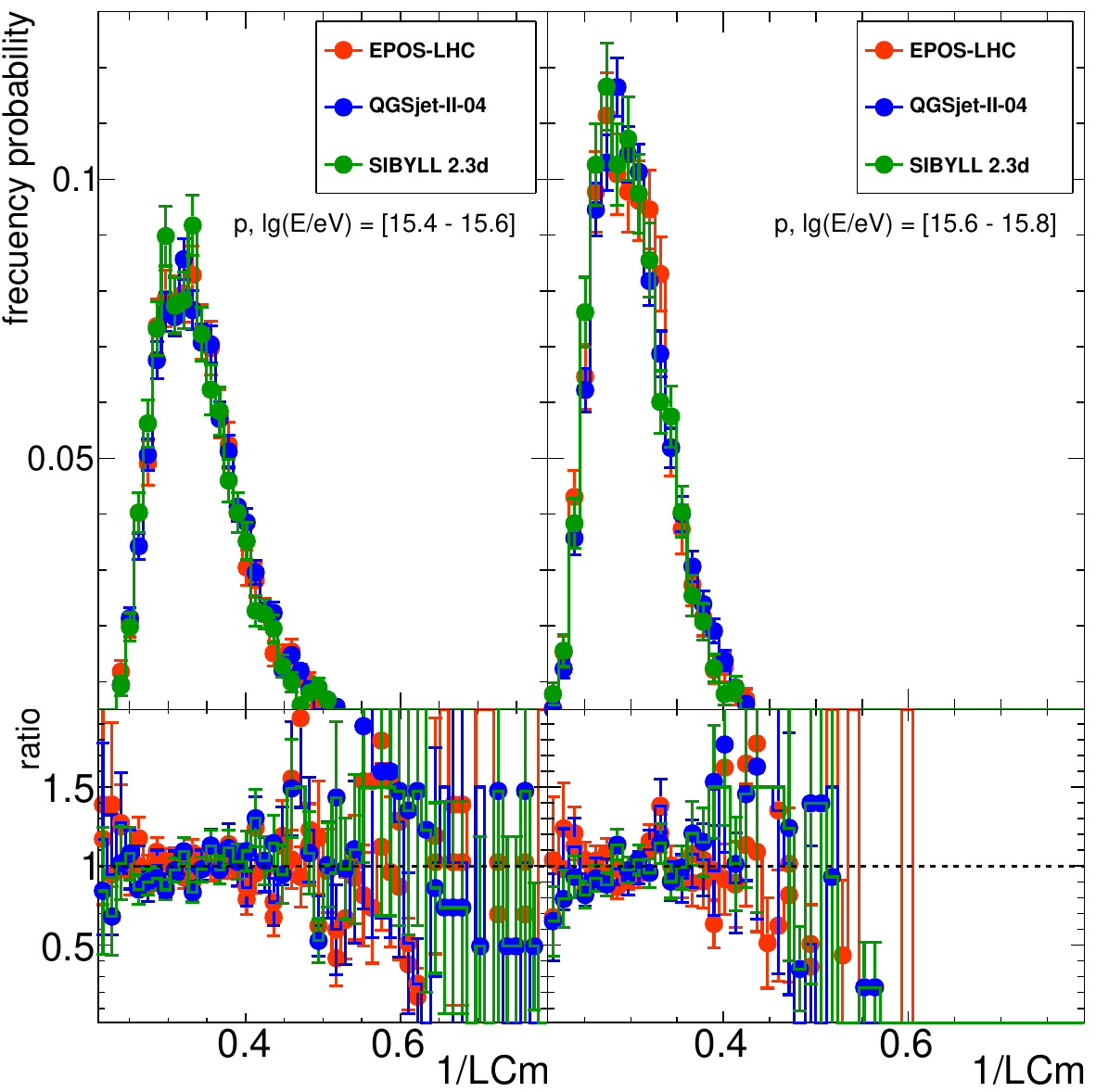}
\hfill
\includegraphics[width=0.49\textwidth]{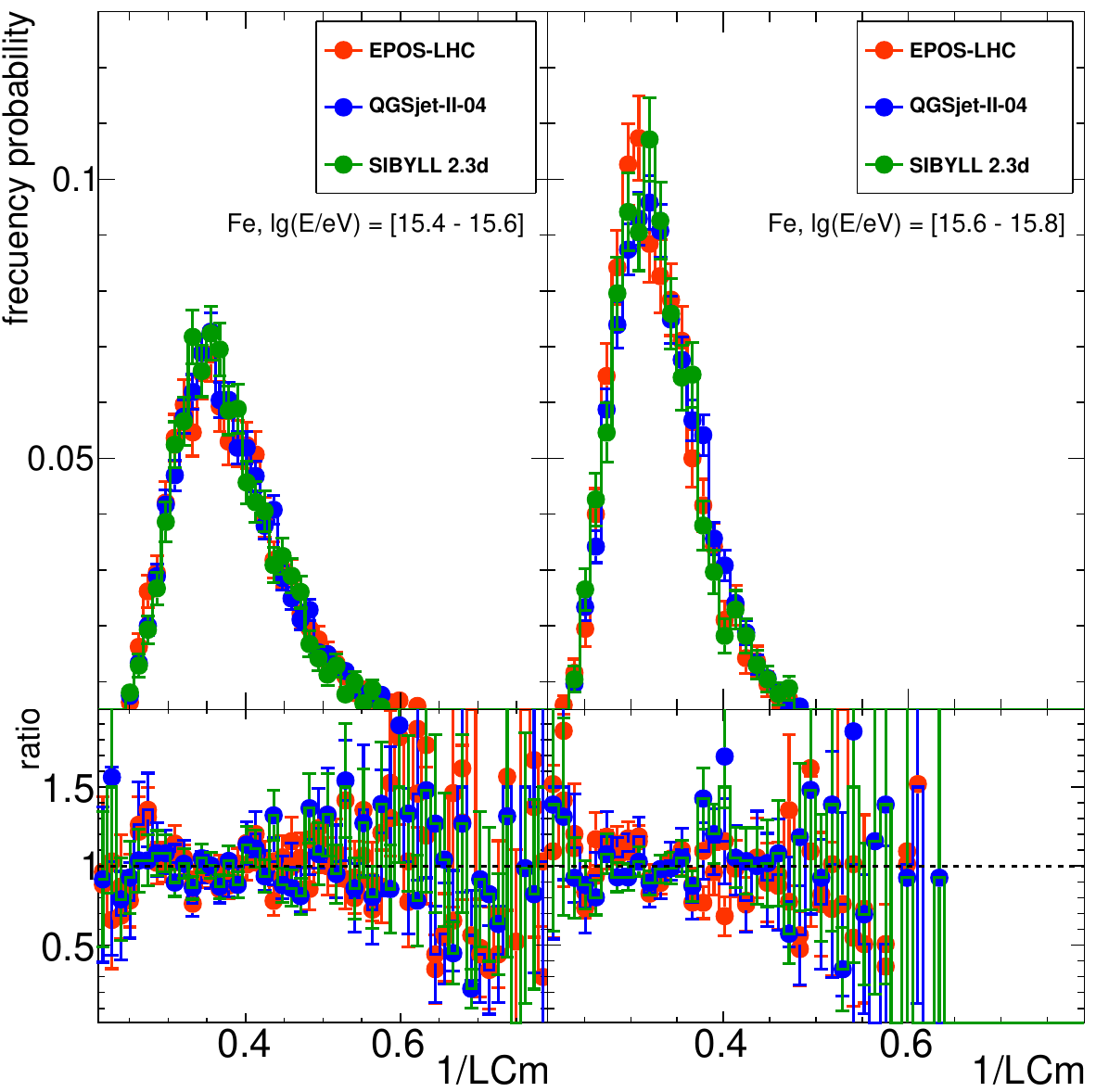}
\caption{The $LCm^{-1}$ distributions in the energy range $\lg(E/\mathrm{eV}) = [15.4 - 15.6]$ and $\lg(E/\mathrm{eV}) = [15.6 - 15.8]$ for proton \textit{(left)} and Fe \textit{(right)} induced showers as predicted by the three hadronic interaction models. Bottom plots display the ratio between each pair of two distributions: EPOS-LHC - QGSjet-II-04 \textit{(red)}, EPOS-LHC - SIBYLL~2.3d \textit{(blue)} and QGSjet-II-04 - SIBYLL~2.3d \textit{(green)}.
\label{fig:LCm_combined}}
\end{figure*}

In Figure \ref{fig:LCm_combined}, we show the distributions of the $LCm^{-1}$ parameter for showers induced by protons \textit{(left)} and iron nuclei \textit{(right)} in two energy intervals, $\lg(E/\mathrm{eV}) = [15.4 - 15.6]$ and $\lg(E/\mathrm{eV}) = [15.6 - 15.8]$, respectively, reconstructed based on full MC simulations of the KASCADE array, considering all three hadronic interaction models: EPOS-LHC, QGSjet-II-04, and SIBYLL 2.3d. The bottom plots show the ratio between each pair of two distributions to illustrate the extent of the differences in predictions among the various models.
We chose to plot the distributions of $LCm^{-1}$, because we adopted the convention that the distributions of parameters sensitive to the primary particle mass, when projected from pattern space to feature space according to Equation \ref{eq:proj}, should be ordered such that the distributions of lighter elements appear to the left of those of heavier elements.

It is well known that iron induced showers produce more muons than proton induced showers of the same energy, due to higher multiplicity in the initial hadronic interactions. At the same time, due to the larger cross-section of heavier nuclei compared to protons, they interact higher in the atmosphere, causing the electromagnetic component to attenuate more significantly. As a result, iron-induced EAS produce fewer electrons at ground level compared to proton-induced showers.

The energy deposited in the $\gamma/e$ detectors or muon detectors is converted into number of particle after applying a Lateral Energy Correction Function (LECF), determined from MC simulations, which removes the contribution of other particle types that produce signals in the $\gamma/e$ and muon detectors, respectively \citet{KASCADE:2000bwl, KASCADE:2005nyh}.
The number of muons $N_\mu$, the number of electrons $N_{e}$, as well as the $age$ parameter ($s$) are obtained based on a modified version of the Nishimura-Kamata-Greisen (NKG) function \citet{KASCADE:2000bwl}:
\begin{equation}\label{NKG}
 \rho_{e,\mu}(r) = C(s) \cdot N_{e,\mu} {\left( r \over r_m^{e,\mu} \right)^{s-\alpha}} {\left( 1 + {r \over r_m^{e,\mu}} \right)^{s-\beta}},
\end{equation}
where
\begin{equation}
C(s) = {\Gamma(\beta - s) \over 2\pi r_m^2 \Gamma(s - \alpha + 2)(\beta - 2s)},
\end{equation}

\noindent

 where $\alpha = 2$, $\beta = 4.5$ and $r_m$ represents the Moli\`ere radius: $r_m = 89$ m and $r_m = 420$ m for electrons and muons, respectively \citet{KASCADE:2000bwl}.
The details of the reconstruction of these three parameters are thoroughly described in \citet{kcdc_manual}.

\begin{figure*}[ht!]
\centering
\includegraphics[width=0.49\textwidth]{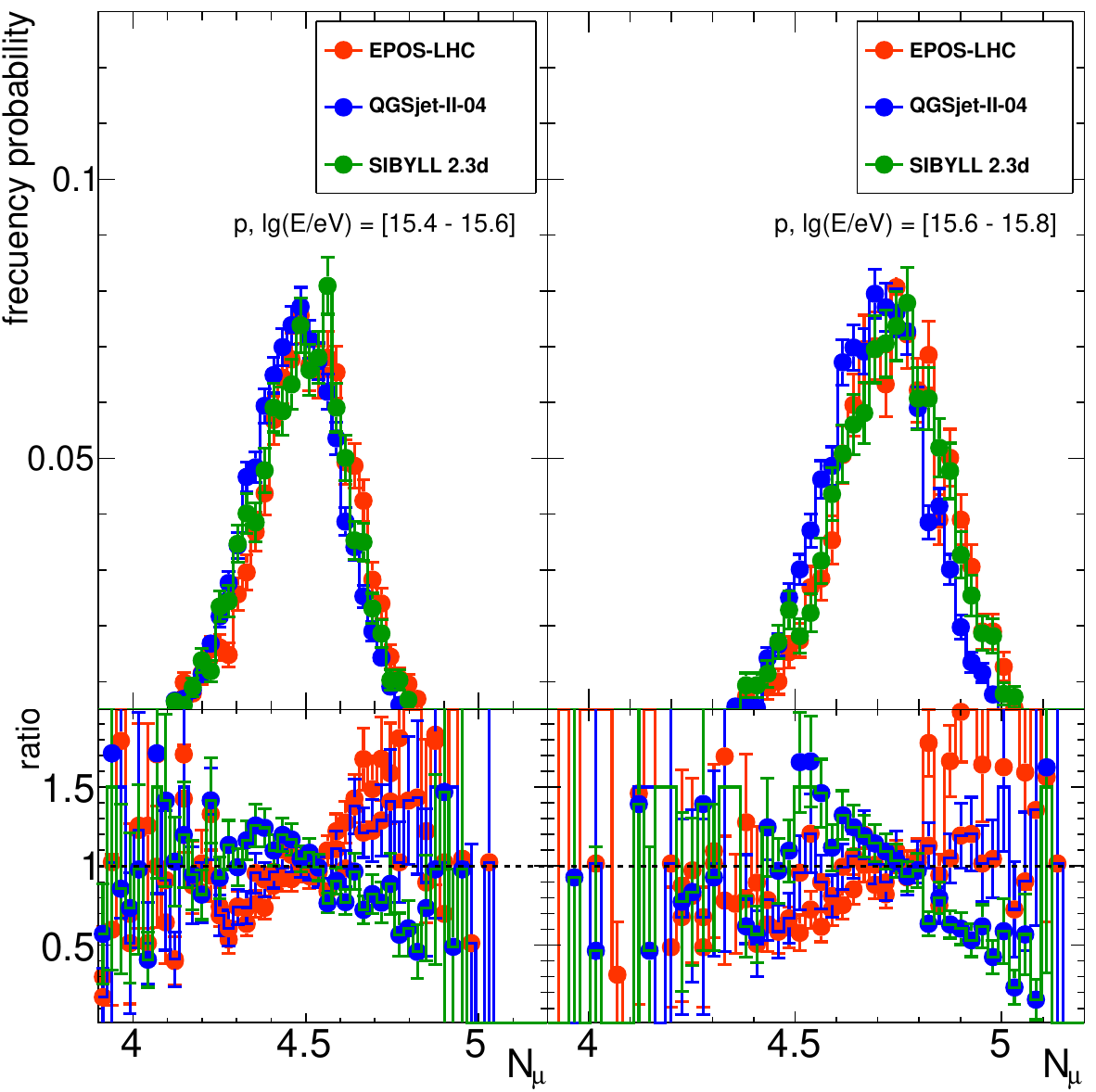}
\hfill
\includegraphics[width=0.49\textwidth]{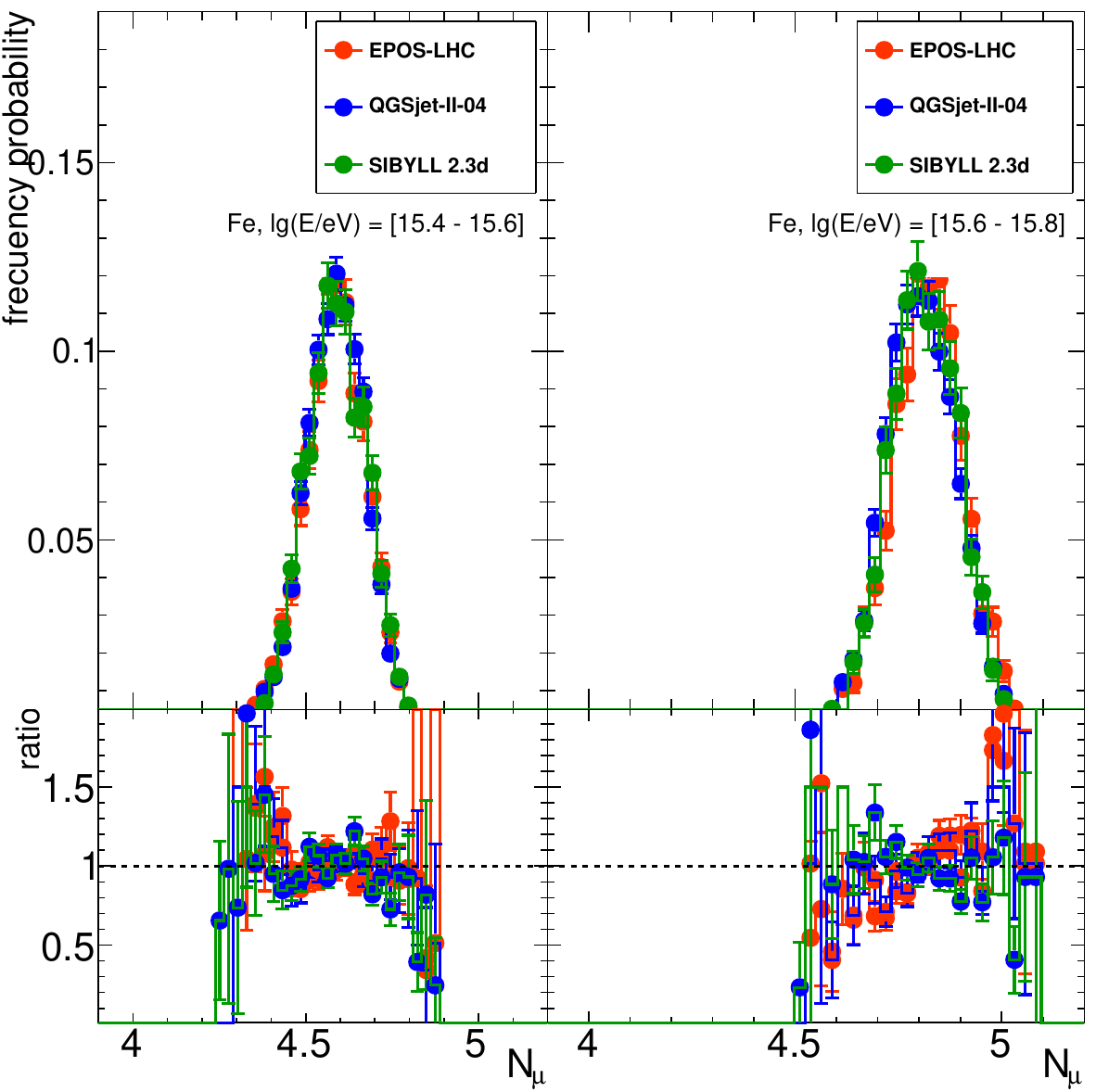}
\caption{The $N_{\mu}$ distributions in the energy range $\lg(E/\mathrm{eV}) = [15.4 - 15.6]$ and $\lg(E/\mathrm{eV}) = [15.6 - 15.8]$ for proton \textit{(left)} and Fe \textit{(right)} induced showers as predicted by the three hadronic interaction models. Bottom plots display the ratio between each pair of two distributions: EPOS-LHC - QGSjet-II-04 \textit{(red)}, EPOS-LHC - SIBYLL~2.3d \textit{(blue)} and QGSjet-II-04 - SIBYLL~2.3d \textit{(green)}.
\label{fig:Nmu_combined}}
\end{figure*}

\begin{figure*}[ht!]
\centering
\includegraphics[width=0.49\textwidth]{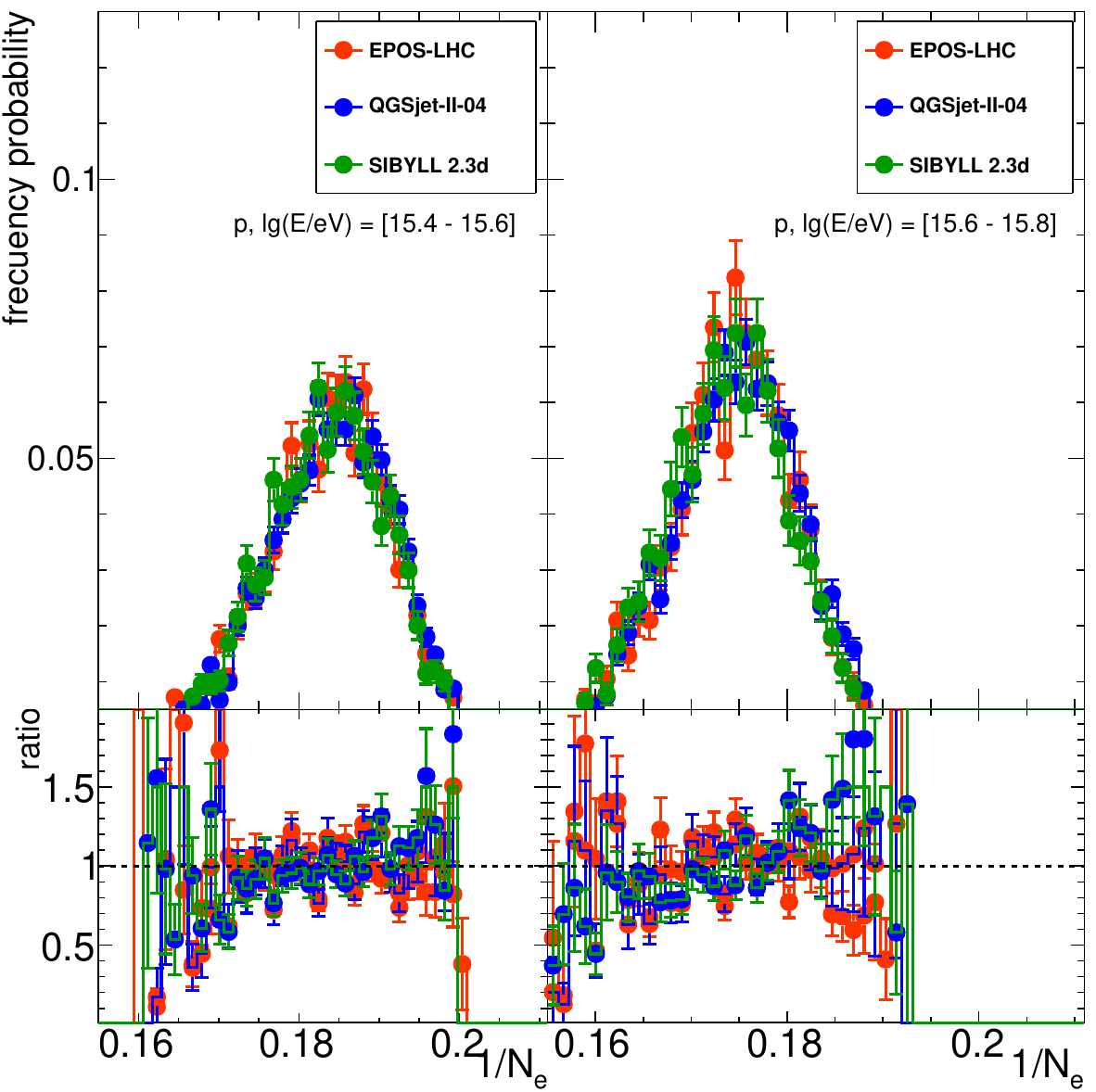}
\hfill
\includegraphics[width=0.49\textwidth]{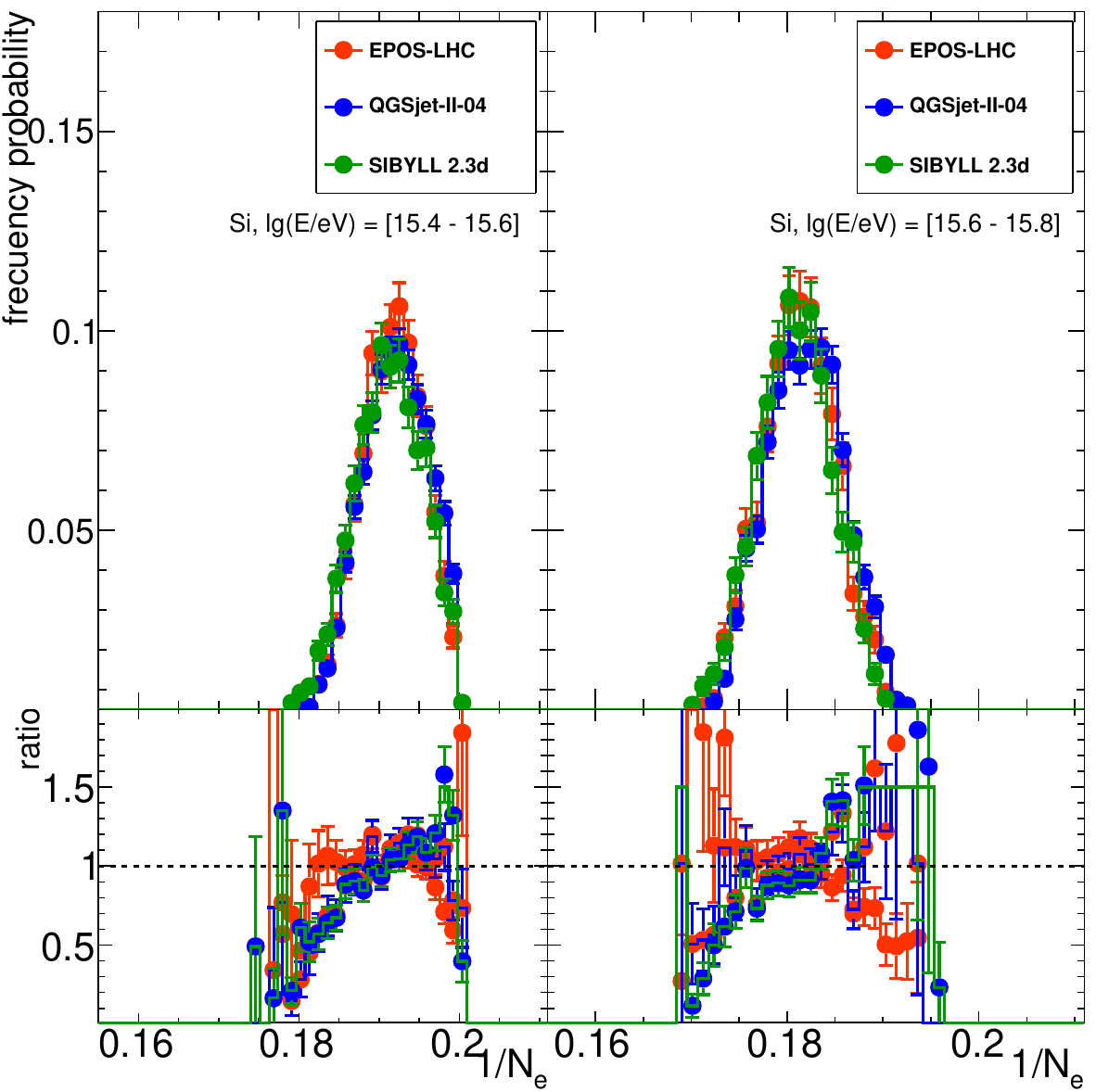}
\caption{The $N_{e}^{-1}$ distributions in the energy range $\lg(E/\mathrm{eV}) = [15.4 - 15.6]$ and $\lg(E/\mathrm{eV}) = [15.6 - 15.8]$ for proton \textit{(left)} and Si \textit{(right)} induced showers as predicted by the three hadronic interaction models.
Bottom plots display the ratio between each pair of two distributions: EPOS-LHC - QGSjet-II-04 \textit{(red)}, EPOS-LHC - SIBYLL~2.3d \textit{(blue)} and QGSjet-II-04 - SIBYLL~2.3d \textit{(green)}.
\label{fig:Ne_combined}}
\end{figure*}

The distributions of the muon number $N_\mu$, obtained from full MC simulations and reconstruction techniques based on the KRETA package, are shown in Figure \ref{fig:Nmu_combined} for EASs induced by protons and iron nuclei in two energy intervals $\lg(E/\mathrm{eV}) = [15.4 - 15.6]$ and $\lg(E/\mathrm{eV}) = [15.6 - 15.8]$, considering the three hadronic interaction models. The values of the muon distribution ratios predicted by the three hadronic interaction models, as shown in the bottom plots, indicate a strong dependence of this observable on the chosen interaction model.
The same dependence on the interaction model can also be observed in the distributions of the electron number $N_e$ in Figure \ref{fig:Ne_combined}. As in the case of the $LCm$ parameter, the electron number distributions are represented as $N_{e}^{-1}$ so that lighter elements appear to the left of heavier ones. Note that the quantities $N_\mu$ and $N_e$ represent the logarithm of the number of muons and electrons, respectively, and this is how we will refer to them hereafter.

\begin{figure*}[ht!]
\centering
\includegraphics[width=0.49\textwidth]{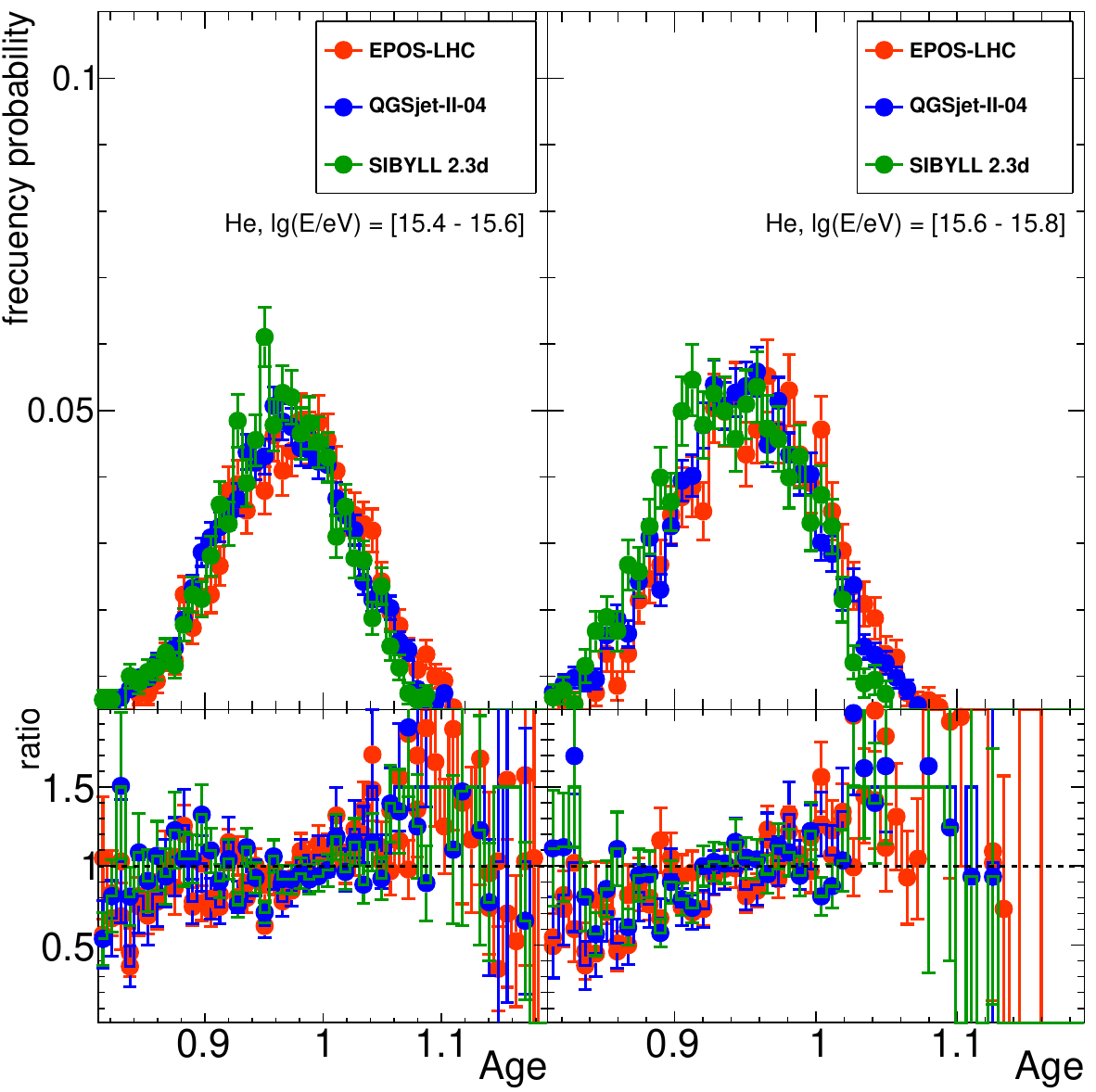}
\hfill
\includegraphics[width=0.49\textwidth]{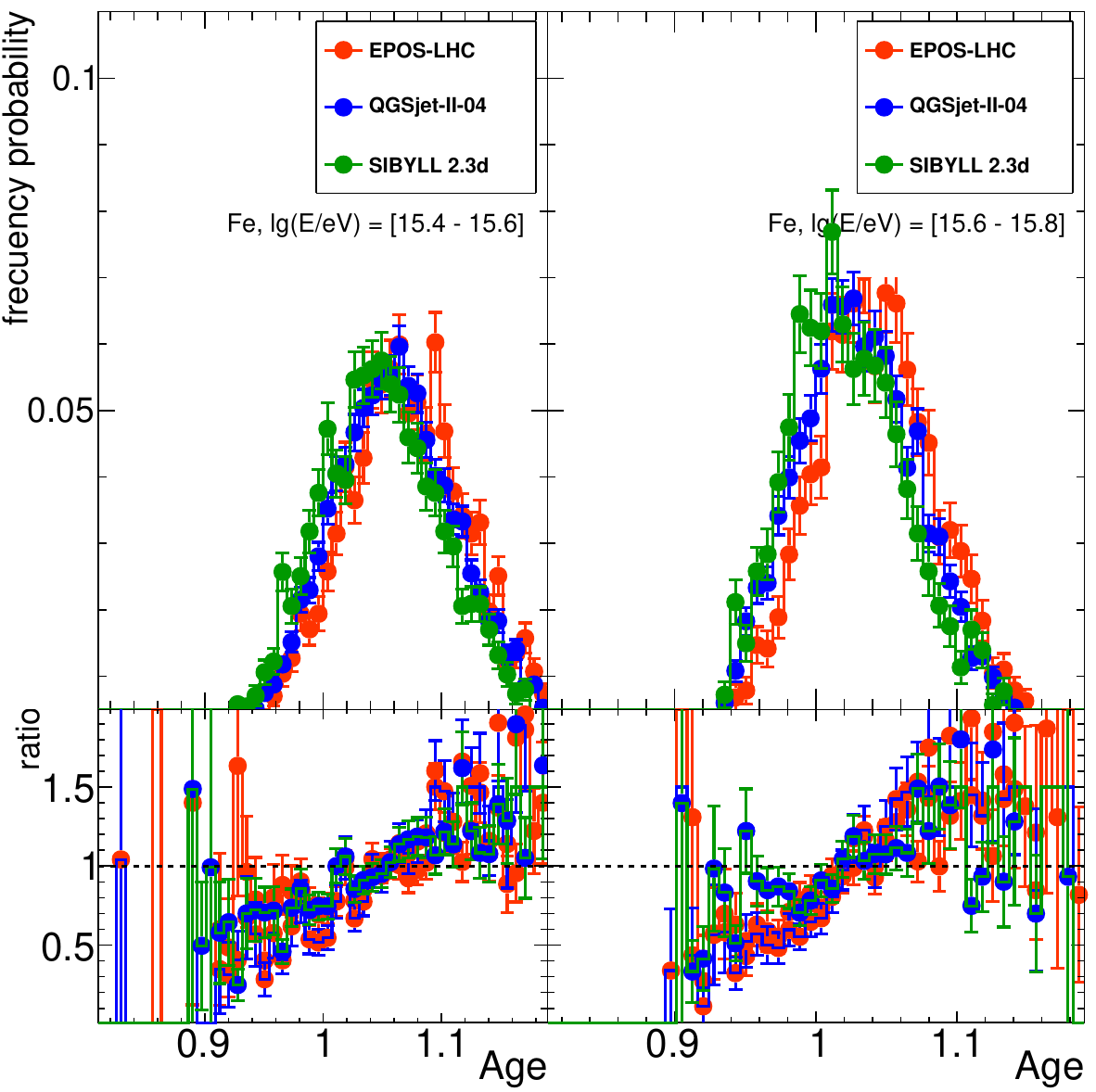}
\caption{The distributions of $Age$ parameter in the energy range $\lg(E/\mathrm{eV}) = [15.4 - 15.6]$ and $\lg(E/\mathrm{eV}) = [15.6 - 15.8]$ for He \textit{(left)} and Fe \textit{(right)} induced showers as predicted by the three hadronic interaction models.
Bottom plots display the ratio between each pair of two distributions: EPOS-LHC - QGSjet-II-04 \textit{(red)}, EPOS-LHC - SIBYLL~2.3d \textit{(blue)} and QGSjet-II-04 - SIBYLL~2.3d \textit{(green)}.
\label{fig:Age_combined}}
\end{figure*}

The $age$ parameter $s$ describes the steepness of the lateral distribution function of electrons, which can vary depending on the specific parameters used in the NKG function. Nevertheless, as shown in Figure \ref{fig:Age_combined}, considering the parameterizations of the NKG function used in the reconstruction of KASCADE data, it is a parameter sensitive to the nature of the primary particle, though strongly dependent on the hadronic interaction model.

To ensure the highest level of quality and consistency in the reconstruction process of both experimental and simulated data, we applied the 'Data Selection Cuts KASCADE' as well as the 'Advised Cuts' \citet{kcdc_manual} recommended and used in the majority of analyses conducted by the KASCADE collaboration. Some of these cuts include: $N_\mu > 2$, $\theta < 42^\circ$, the $age$ parameter $s = [0.6 - 1.3]$ and $N_e > 5$. It is worth highlighting that the trigger efficiency reaches 100\% for showers with $N_e > 4.25$, which corresponds to a primary energy of approximately $\lg(E/\mathrm{eV}) \simeq 14.8$, thus covering the energy range of interest in this study, $\lg(E/\mathrm{eV}) > 15.0$.

Next, we use the simulated distributions of the four parameters sensitive to the nature of the primary particle, for all five primary species, and apply the PCA technique described in Section \ref{sec:pca} to project these values from \textit{pattern} space to \textit{feature} space, with the aim of optimally capturing the variance in the data and thereby enhancing the separation between species. It is worth noting that the MC events were binned according to the true primary energy from CORSIKA, whereas for the experimental data, energy binning was performed by taking into account the detector resolution and bin-to-bin migration effects, as estimated from MC simulations.

\section{Mass composition around the \textit{knee} using PCA} \label{sec:mass_comp}

The input values considered in the PCA analysis (see Section \ref{sec:pca}) are represented by our set of five primary particles $M$ = $\{$p, He, C, Si, Fe$\}$, while the set of $P$ observables is given by the four parameters obtained from simulations:
\begin{equation}
\begin{aligned}
x_0 &= LCm^{-1} \\
x_1 &= N_{\mu} \\
x_2 &= N_e^{-1} \\
x_3 &= age.
\end{aligned}
\end{equation}
For each energy interval of width $\lg(E/\mathrm{eV}) = 0.2$ in the range $\lg(E/\mathrm{eV}) = [15.0 - 16.0]$ and each hadronic interaction model, we perform the projection from \textit{pattern} space to \textit{feature} space. We sort the eigenvalues $\lambda_n$ of the covariance matrix in descending order and select the first two principal components (PCA0 and PCA1), corresponding to the eigenvectors associated with the two largest eigenvalues.

\begin{figure*}[ht!]
\centering
\includegraphics[width=0.49\textwidth]{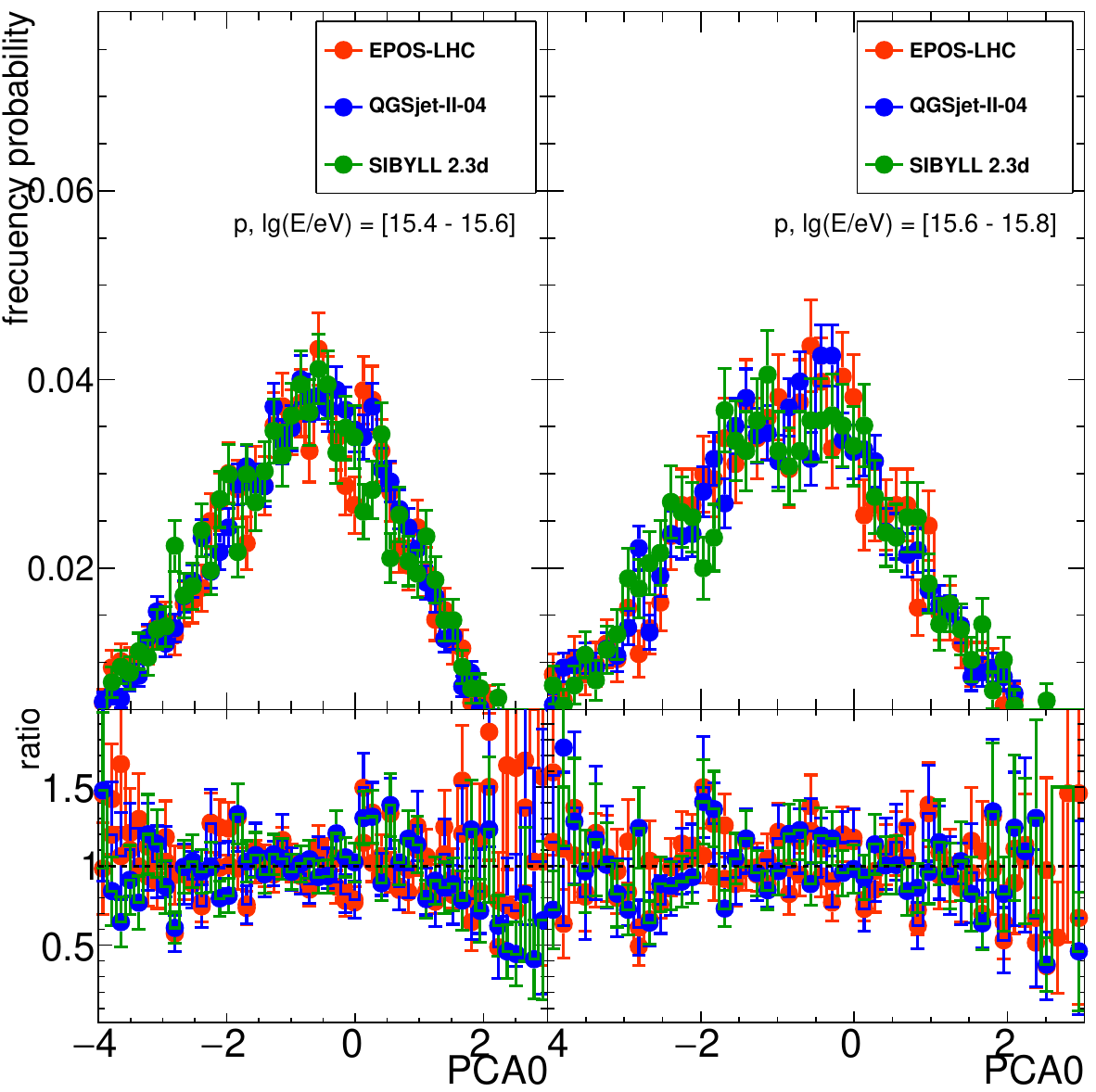}
\hfill
\includegraphics[width=0.49\textwidth]{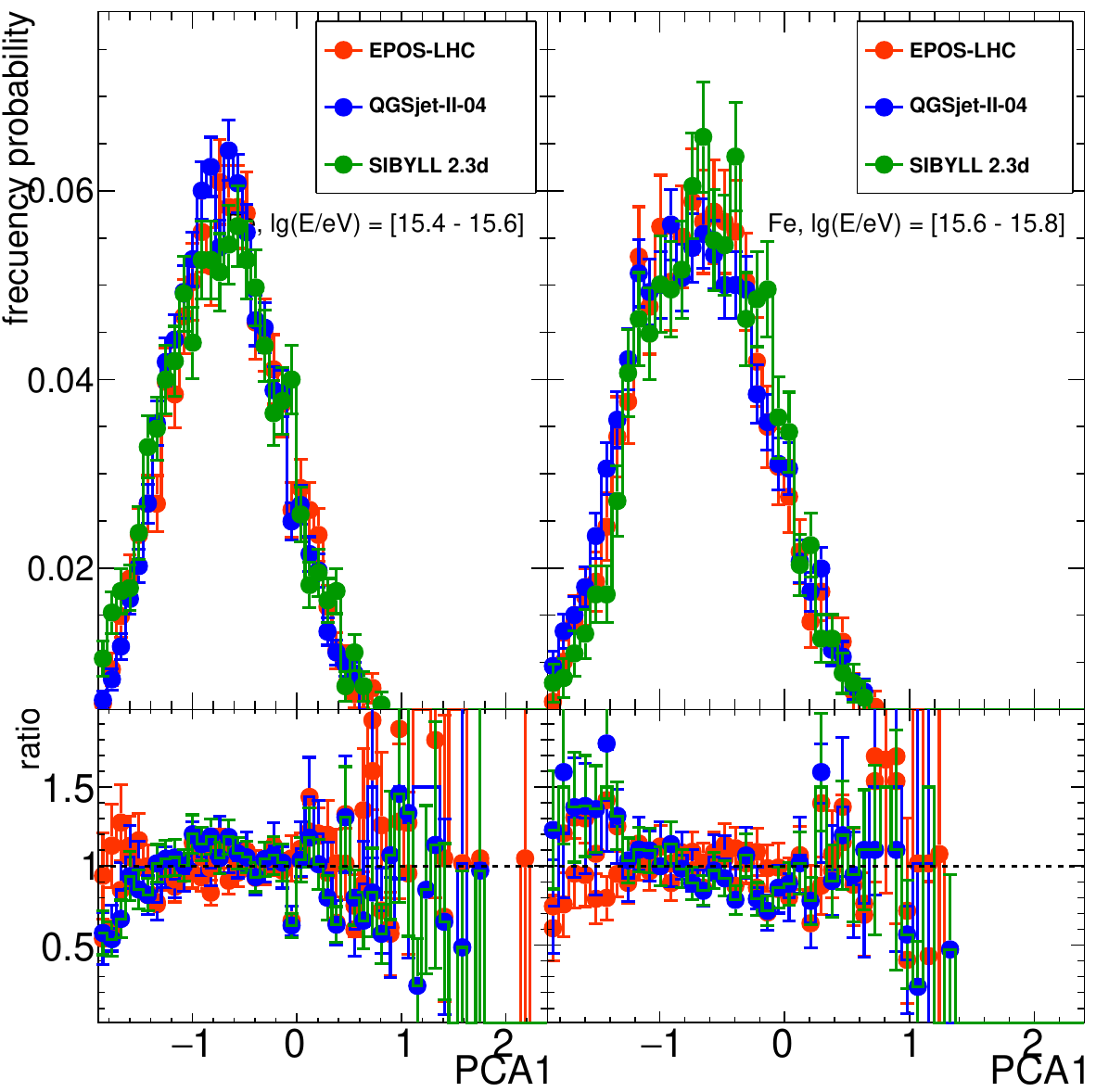}
\caption{The distributions of PCA0 for proton induced showers \textit{(left)} and PCA1 for iron induced showers \textit{(right)} in the energy range $\lg(E/\mathrm{eV}) = [15.4 - 15.6]$ and $\lg(E/\mathrm{eV}) = [15.6 - 15.8]$ for three hadronic interaction models. Bottom plots display the ratio between each pair of two distributions: EPOS-LHC - QGSjet-II-04 \textit{(red)}, EPOS-LHC - SIBYLL~2.3d \textit{(blue)} and QGSjet-II-04 - SIBYLL~2.3d \textit{(green)}.
\label{pca0pca1}}
\end{figure*}

In Figure \ref{pca0pca1}, we present the one-dimensional distributions of PCA0 values for proton-induced showers, and the PCA1 distributions for iron-induced showers, in two energy intervals, $\lg(E/\mathrm{eV}) = [15.4 - 15.6]$ and $\lg(E/\mathrm{eV}) = [15.6 - 15.8]$, based on the three hadronic interaction models. 
It can be observed that the PCA0 and PCA1 distributions exhibit a remarkable level of agreement among the three hadronic interaction models used. Such a result is not entirely unexpected, considering that the parameter $LCm$ is nearly independent of the hadronic model considered, while the ratio $N_\mu / N_e$ within this energy range has been shown to exhibit minimal sensitivity to the chosen interaction model \citet{Tian:2024vor}.

\begin{figure}[ht!]
\includegraphics[width=0.49\textwidth]{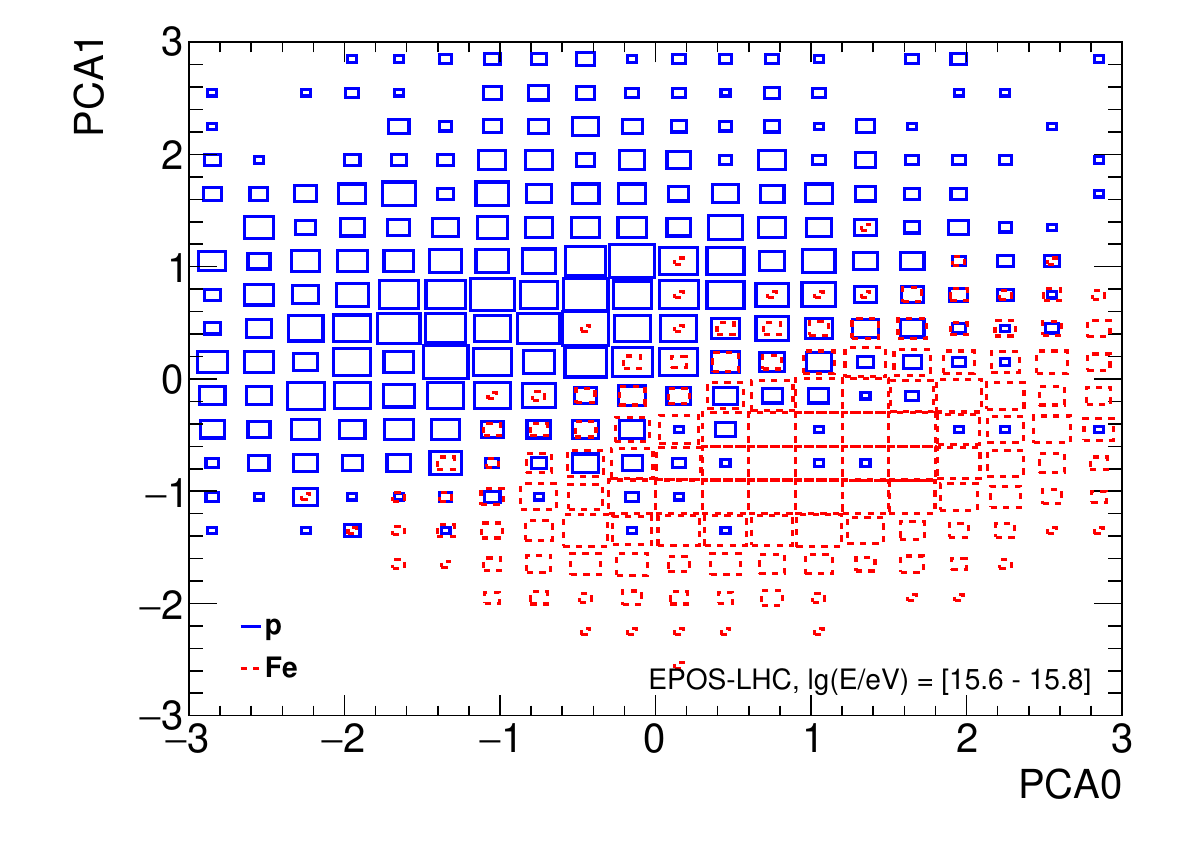}
\caption{The two-dimensional PCA0 vs. PCA1 distributions for proton and iron induced showers in the energy interval $\lg(E/\mathrm{eV}) = [15.6 - 15.8]$ based on the EPOS-LHC model. The size of the squares reflects the number of events in each bin.
\label{fig:2Dpca01}}
\end{figure}

Figure \ref{fig:2Dpca01} presents the two-dimensional PCA0 vs. PCA1 distribution for air showers induced by protons and iron nuclei within the energy interval $\lg(E/\mathrm{eV}) = [15.6 - 15.8]$, based on simulations using the EPOS-LHC hadronic interaction model. As can be seen from these distributions, PCA0 captures the largest amount of variance in the data.

As described in Section \ref{sec:pca}, retaining only the first two principal components preserves most of the information related to the separability between different primary species, at the cost of losing the remaining information contained in the last two principal components, which are associated with the smallest eigenvalues. We quantified the separability between proton-induced and iron-induced events using the Figure of Merit (FOM) as follows: we performed a Fischer projection of the two-dimensional PCA0 vs. PCA1 distributions onto a one-dimensional distribution and calculated the FOM for proton- and iron-induced events

\begin{equation}
 FOM = \frac {|\mu_p - \mu_{Fe}|} {\sqrt{\sigma_p^2 + \sigma_{Fe}^2}},
\end{equation}
where $\mu$ and $\sigma$ denote the mean and standard deviation of the distributions.
In almost all energy intervals, we obtained FOM values greater than 2. In comparison, the separability between proton and iron events based solely on individual classical observables such as $N_\mu$, $N_e$, $LCm$, and $Age$ parameter yields FOM values around 1.

\begin{figure*}[ht!]
\centering
\includegraphics[width=0.99\textwidth]{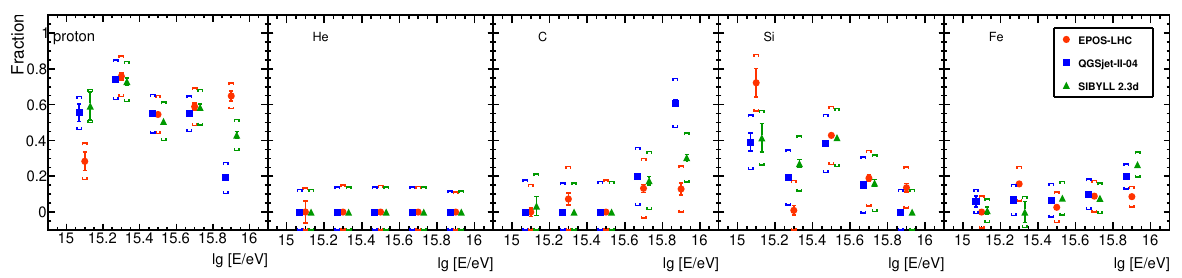}
\caption{The evolution of the individual fractions of the five primary particle species as a function of energy, obtained based on the three hadronic interaction models. The inner error bars represent the statistical uncertainties, while the outer error brackets represent the systematic uncertainties (see Section \ref{sec:syst}). The X-axis value of each point corresponds to the center of the energy bin; however, to improve visual clarity, they have been artificially shifted along the X-axis.
\label{fig:fractions}}
\end{figure*}

Using the same reconstruction procedures, we applied the PCA method to the KASCADE experimental data, thereby constructing two-dimensional distributions of PCA0 vs. PCA1 for the five energy intervals within the $\lg(E/\mathrm{eV}) = [15.0 - 16.0]$ range. Subsequently, we fitted these experimental 2D distributions with the 2D distributions obtained from MC simulations for the five primary particle species, for each hadronic interaction model, following a Chi-squared minimization. In this way, we obtain the abundance of each primary particle species as a function of energy within the studied energy range. Figure \ref{fig:fractions} shows the evolution of the individual fractions for the five types of primary particles (p, He, C, Si, and Fe) as a function of energy, based on the fits of the experimental PCA0 vs. PCA1 distributions to the MC predictions corresponding to the three considered hadronic interaction models. 
It is worth mentioning that the concentrations of the different primary species as a function of energy obtained by this method, based on the four EAS parameters, do not exhibit any discrepancy between the hadronic models considered in the simulation process.
Another important point is that the fractions of protons and iron nuclei (the two extremes in the sets of nuclei considered in the analyses) obtained using this method based on KASCADE data are in excellent agreement with those obtained by the IceTop experiment at energies above the \textit{knee}, whereas the helium nuclei fractions show only a slight overlap within the systematic uncertainty bands \citet{Rawlins_2016, IceCube:2019hmk, Soldin:2022evu}. A quantitative comparison of the intermediate nuclei abundances would be difficult, given that the IceTop analyses considered only four elements: p, He, O, and Fe, while in the present analysis we used a set of five elements: p, He, C, Si, and Fe.
The details of the systematic uncertainty analysis are presented in Section \ref{sec:syst}. 

\begin{figure}[ht!]
\includegraphics[width=0.49\textwidth]{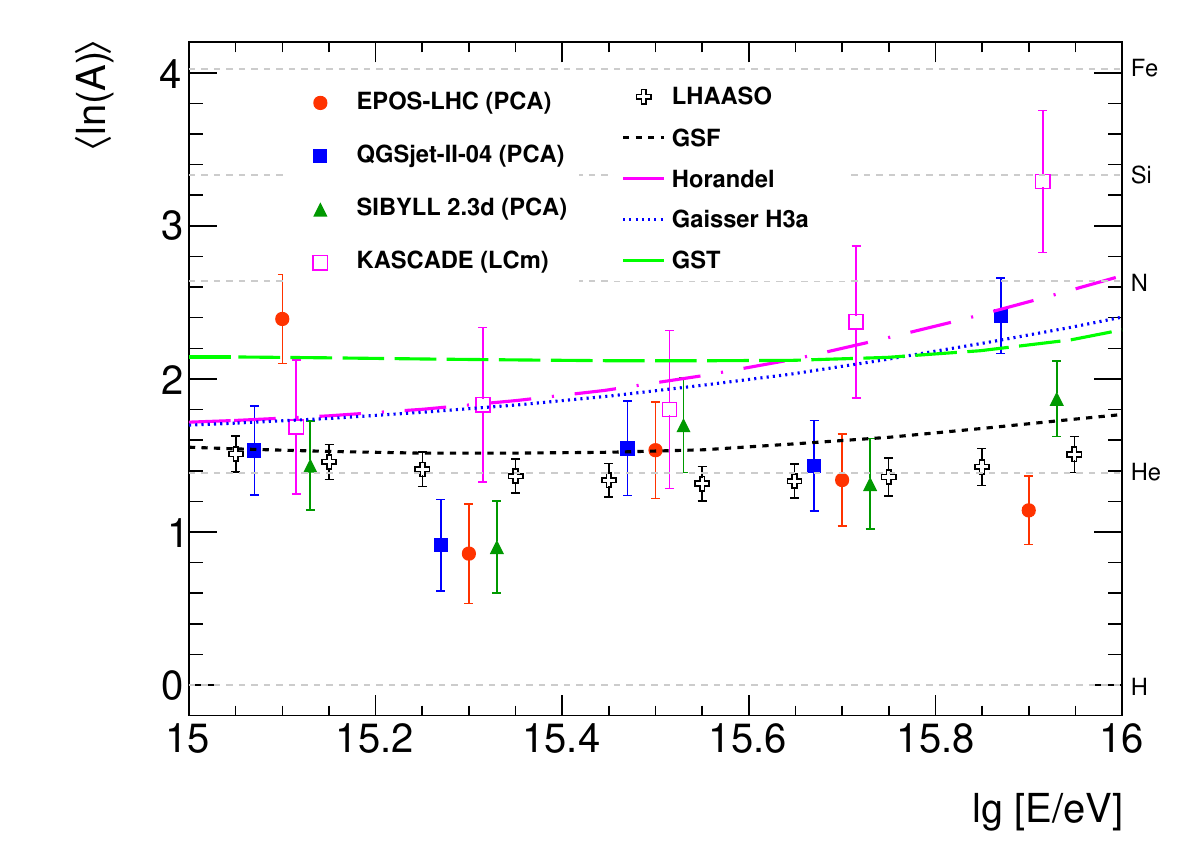}
\caption{The evolution of $\langle \ln (A) \rangle$ as a function of energy, obtained using the PCA method based on the three hadronic interaction models. For comparison, we also plotted the evolution of $\langle \ln (A) \rangle$ as a function of energy derived solely from the $LCm$ parameter based on the KASCADE experimental data \citet{2023JCAP...09..020A}, as well as recent results from the LHAASO$-$KM2A experiment \citet{LHAASO:2024knt}. These results are further compared with the predictions of four data-driven and astrophysical models: GSF \citet{Dembinski:2017zsh}, Horandel \citet{Hoerandel:2002yg}, Gaisser H3a \citet{Gaisser:2013bla} and GST \citet{STANEV201442}.
\label{fig:lnA}}
\end{figure}

Next, we converted the relative abundances of the different primary species as a function of energy into $\langle \ln (A) \rangle$ units, where $A$ represents the mass number of each primary nuclei. In Figure \ref{fig:lnA}, we presented the evolution of $\langle \ln (A) \rangle$ as a function of energy based on the results obtained using this PCA method applied to the KASCADE experimental data, using three hadronic interaction models. We also included for comparison the results previously obtained solely based on the $LCm$ parameter extracted from the KASCADE data \citet{2023JCAP...09..020A}, as well as the recent results from the LHAASO$-$KM2A experiment \citet{LHAASO:2024knt}. Superimposed on these experimental results are the theoretical predictions of the evolution of the mass composition $\langle \ln (A) \rangle$ as a function of energy around the \textit{knee} from four data-driven and astrophysical models: GSF \citet{Dembinski:2017zsh}, Horandel \citet{Hoerandel:2002yg}, Gaisser H3a \citet{Gaisser:2013bla} and GST \citet{STANEV201442}.

The values of the mean logarithmic mass $\langle \ln (A) \rangle$ together with the associated uncertainties obtained in this work are listed in Table \ref{tab:lnA} for each hadronic interaction model.

\begin{table}[h!]
\centering
\small
\setlength{\tabcolsep}{6pt}
\begin{tabular}{c c c c}
\hline
$\lg(E/\rm{eV})$ & EPOS-LHC & QGSjet-II-04 & SIBYLL 2.3d \\
\hline
15.10 & 2.39 $\pm$ 0.29 & 1.53 $\pm$ 0.29 & 1.43 $\pm$ 0.29 \\
15.30 & 0.86 $\pm$ 0.33 & 0.91 $\pm$ 0.30 & 0.90 $\pm$ 0.30 \\
15.50 & 1.53 $\pm$ 0.32 & 1.55 $\pm$ 0.31 & 1.70 $\pm$ 0.31 \\
15.70 & 1.34 $\pm$ 0.30 & 1.43 $\pm$ 0.30 & 1.31 $\pm$ 0.30 \\
15.90 & 1.14 $\pm$ 0.22 & 2.41 $\pm$ 0.25 & 1.87 $\pm$ 0.25 \\
\hline
\end{tabular}
\caption{The values of $\langle \ln(A) \rangle$ and associated uncertainties as a function of energy for each hadronic interaction model.}
\label{tab:lnA}
\end{table}

As shown in Figure \ref{fig:lnA}, the results obtained in this work are in excellent agreement with the results of the LHAASO$-$KM2A experiment around the \textit{knee} within the limits of systematic uncertainties. It is worth mentioning that the results of the LHAASO$-$KM2A experiment are based on the reconstruction of the electromagnetic component and the number of muons with very high precision, along with the reconstruction of the primary energy in a way that is independent of the mass composition and hadronic interaction model. When comparing to the predictions of data-driven and astrophysical models, we observe that the GSF model most accurately describes the mass composition around the \textit{knee}, as it fits both the experimental data obtained in this work using the PCA method and the results from LHAASO–KM2A.

 The difference found between our results and the previous KASCADE results \citet{2005APh....24....1A, 2013arXiv1306.6283T} could be explained by the use of the updated hadronic interaction models. Particularly, it is worth mentioning that our result for $\langle \ln (A) \rangle$ as a function of energy obtained in this analysis is in agreement with the results presented in \citet{Kuznetsov:2023pvo} based on KASCADE data reconstructed using a novel machine learning technique.

\subsection{Systematic uncertainties} \label{sec:syst}

We reconstruct the mass composition by fitting KASCADE experimental data (2D distributions of PCA0 vs. PCA1) with MC templates in each energy bin, accounting for systematic uncertainties from primary energy reconstruction and simulation dependencies.

Systematic errors in energy reconstruction were estimated by comparing true energies from \textsc{CORSIKA} with reconstructed energies obtained using the CRES and KRETA simulation and reconstruction chain. The relative energy difference \((E^{\mathrm{true}} - E^{\mathrm{rec}})/E^{\mathrm{true}}\) was evaluated in small energy bins, averaged over three hadronic interaction models and five primary species. Biases (defined as the mean of the relative difference between true and reconstructed
energy) were found between 1\%--3\%, and systematic uncertainties (i.e., energy resolution) ranged from 20\%--29\%.

To reduce correlations between adjacent bins, we used wider energy intervals of \(\lg(E/\mathrm{eV}) = 0.2\). We also corrected for bin-to-bin migration by re-binning the data based on model-dependent migration probabilities derived from simulations.
Since migration effects vary slightly with the type of primary particle, we tested multiple composition scenarios. The best fit was obtained assuming an equal mix of light and heavy nuclei, though the results were not significantly affected when assuming either a light- or heavy-dominated composition.

Next, we tested the sensitivity of the method and estimated the bias and systematic errors of the reconstructed fractions due to MC dependencies. In the first step, we considered mock data sets (2D distributions of PCA0 vs. PCA1) generated from the predictions of a hadronic interaction model, including random but known concentrations of the five types of primary particles (p, He, C, Si, and Fe), and fitted them using distributions obtained from MC simulations based on the same interaction model. By repeating this process a sufficiently large number of times, we estimated the first set of biases and systematic errors—those arising from the sensitivity of the method itself—considering the $68\%$ confidence contours of the distributions of the reconstructed and true fractions, denoted as $\sigma_1$.

We repeated the same procedure, this time using mock data sets generated based on the predictions of one hadronic interaction model, and performed the reconstruction by fitting these distributions with MC predictions from a different interaction model. This was repeated for all combinations of models considered in this study and we obtained in this way the second source of biases and systematic errors $\sigma_2$ due to MC mismodeling.

While the bias values are very close to zero, the systematic errors of the individual fractions, $\sigma_2$, were consistently larger than $\sigma_1$. Therefore, we chose to apply only the second set of systematic errors in the reconstruction process from experimental data, as shown in Figure \ref{fig:fractions} and propagated in Figure \ref{fig:lnA}, using the general error propagation formula, applied to the primary fractions and their covariance matrix, including the correlations. The values of the individual fractions of different species together with the systematic errors ($\sigma_2$) as a function of energy for each hadronic interaction model are listed in Table \ref{tab:fraction_errors}.

\begin{table*}[htbp]
\centering
\small
\setlength{\tabcolsep}{5pt}
\begin{tabular}{c c c c c}
\hline
$\lg(E/\rm{eV})$ & Particle & EPOS-LHC (Frac. $\pm$ $\sigma_2$) & QGSjet-II-04 (Frac. $\pm$ $\sigma_2$) & SIBYLL 2.3d (Frac. $\pm$ $\sigma_2$) \\ 
\hline
15.10 & p & 0.28 $\pm$ 0.10 & 0.55 $\pm$ 0.09 & 0.59 $\pm$ 0.09 \\ 
15.30 & p & 0.76 $\pm$ 0.11 & 0.74 $\pm$ 0.11 & 0.73 $\pm$ 0.11 \\ 
15.50 & p & 0.55 $\pm$ 0.11 & 0.55 $\pm$ 0.11 & 0.51 $\pm$ 0.11 \\ 
15.70 & p & 0.59 $\pm$ 0.10 & 0.55 $\pm$ 0.10 & 0.59 $\pm$ 0.10 \\ 
15.90 & p & 0.65 $\pm$ 0.07 & 0.19 $\pm$ 0.09 & 0.43 $\pm$ 0.09 \\ 
15.10 & He & 0.00 $\pm$ 0.13 & 0.00 $\pm$ 0.13 & 0.00 $\pm$ 0.13 \\ 
15.30 & He & 0.00 $\pm$ 0.15 & 0.00 $\pm$ 0.14 & 0.00 $\pm$ 0.14 \\ 
15.50 & He & 0.00 $\pm$ 0.15 & 0.00 $\pm$ 0.14 & 0.00 $\pm$ 0.14 \\ 
15.70 & He & 0.00 $\pm$ 0.14 & 0.00 $\pm$ 0.14 & 0.00 $\pm$ 0.14 \\ 
15.90 & He & 0.00 $\pm$ 0.11 & 0.00 $\pm$ 0.12 & 0.00 $\pm$ 0.12 \\ 
15.10 & C & 0.00 $\pm$ 0.15 & 0.00 $\pm$ 0.18 & 0.03 $\pm$ 0.18 \\ 
15.30 & C & 0.07 $\pm$ 0.18 & 0.00 $\pm$ 0.16 & 0.00 $\pm$ 0.16 \\ 
15.50 & C & 0.00 $\pm$ 0.18 & 0.00 $\pm$ 0.17 & 0.00 $\pm$ 0.17 \\ 
15.70 & C & 0.13 $\pm$ 0.17 & 0.20 $\pm$ 0.16 & 0.18 $\pm$ 0.16 \\ 
15.90 & C & 0.13 $\pm$ 0.13 & 0.61 $\pm$ 0.14 & 0.30 $\pm$ 0.14 \\ 
15.10 & Si & 0.72 $\pm$ 0.16 & 0.39 $\pm$ 0.15 & 0.41 $\pm$ 0.15 \\ 
15.30 & Si & 0.01 $\pm$ 0.17 & 0.19 $\pm$ 0.15 & 0.27 $\pm$ 0.15 \\ 
15.50 & Si & 0.43 $\pm$ 0.16 & 0.38 $\pm$ 0.16 & 0.42 $\pm$ 0.16 \\ 
15.70 & Si & 0.19 $\pm$ 0.16 & 0.15 $\pm$ 0.16 & 0.16 $\pm$ 0.16 \\ 
15.90 & Si & 0.14 $\pm$ 0.11 & 0.00 $\pm$ 0.13 & 0.00 $\pm$ 0.13 \\ 
15.10 & Fe & 0.00 $\pm$ 0.09 & 0.06 $\pm$ 0.07 & 0.01 $\pm$ 0.07 \\ 
15.30 & Fe & 0.16 $\pm$ 0.10 & 0.07 $\pm$ 0.09 & 0.00 $\pm$ 0.09 \\ 
15.50 & Fe & 0.03 $\pm$ 0.09 & 0.07 $\pm$ 0.09 & 0.08 $\pm$ 0.09 \\ 
15.70 & Fe & 0.09 $\pm$ 0.09 & 0.10 $\pm$ 0.08 & 0.08 $\pm$ 0.08 \\ 
15.90 & Fe & 0.09 $\pm$ 0.06 & 0.20 $\pm$ 0.07 & 0.27 $\pm$ 0.07 \\ 

\hline
\end{tabular}
\caption{The fraction and systematic uncertainties $\sigma_2$ per particle species and energy for each interaction model (see text).}
\label{tab:fraction_errors}
\end{table*}

\section{Conclusions} \label{sec:conclusions}

In this paper, we applied the PCA method in a multivariate analysis to reconstruct the mass composition of cosmic rays around the \textit{knee}, using experimental data recorded by the KASCADE experiment. We used four EAS parameters that are sensitive to the nature of the primary particle ($LCm$, $N_\mu$, $N_e$, and $age$). Based on full MC simulations of the KASCADE array, we demonstrated that the PCA technique identifies a set of orthogonal axes that best represent the variance in the dataset, enhancing the separation of different primary species.

We fitted the experimental distributions of the first two principal components (PCA0 vs.\ PCA1) with MC predictions for five primary particle species (p, He, C, Si, and Fe) in the energy range $\lg(E/\rm{eV}) = [15 - 16]$. We used three hadronic interaction models (EPOS-LHC, QGSjet-II-04, and SIBYLL 2.3d) for the simulations. We found that, based on the PCA technique, the mass composition results are nearly independent of the interaction model used in the simulation process, although the individual parameters employed remain model dependent.

These results confirm the widely accepted and experimentally validated scenario: around the \textit{knee}, the light component (mainly protons) decreases in abundance, while the heavier components show a slight increase.

The evolution of the mean logarithmic mass $\langle \ln (A) \rangle$ as a function of energy, obtained in this work based on KASCADE data, is in very good agreement with recent results from the LHAASO–KM2A experiment and with the predictions of the GSF model around the \textit{knee}. 
The consistency of our results with the GSF model suggests that the PCA method, applied to EAS parameters extracted from KASCADE data using modern hadronic interaction models, reliably captures the global trends in cosmic-ray composition. The inferred mass composition is derived from a multidimensional description of shower development, incorporating several key observables, which ensures a physically meaningful and robust interpretation.
Therefore, the results presented in this study may serve as an additional reference for reconstructing the mass composition of cosmic rays in the \textit{knee} region, in agreement with the global GSF estimates.

\begin{acknowledgments}
I would like to thank the KCDC team for making the KASCADE experiment data and simulations accessible. I wish to express my gratitude to Octavian Sima for the numerous fruitful discussions and helpful comments that greatly contributed to this work. Special thanks go to my newborn son Darius-Emil, whose arrival right after submission gave me the best motivation to finish this paper.
This work was supported by a grant of the Ministry of Research, Innovation and Digitization, CNCS - UEFISCDI, project number PN-IV-P2-2.1-TE-2023-0069, within PNCDI IV and by the Romanian Ministry of Research, Innovation and Digitalization under the Romanian National Core Program LAPLAS VII-contract no. 30N/2023.
\end{acknowledgments}


\bibliography{references}{}
\bibliographystyle{aasjournalv7}

\end{document}